\documentclass[a4paper]{EslabStyle}
\usepackage{epsfig}

\begin{document}

\title{Observational Constraints on the Theory of the IMF}

\author{B. Elmegreen\inst{1} 
  } \institute{IBM T.J. Watson Research Center, PO Box 218, 
  Yorktown Hts, NY 10598, USA, bge@watson.ibm.com}

\maketitle 

\begin{abstract}

Observational constraints on the theory of the IMF are reviewed. These
observations include the new result that star formation is very rapid,
usually going from start to finish in only 1 or 2 dynamical times on a
wide range of scales. This result, combined with the observation that
the IMF is independent of cluster density over a factor of several
hundred, implies that protostar coagulation during orbital motions in
a cloud are not important for the IMF: there is not enough time. The
observation that the IMF in individual clusters is about the same
as the IMF in whole galaxies implies that stars of all masses form
randomly in self-gravitating clouds of all masses. There cannot be a
sequence of stellar masses in a cluster, based on stirring or heating
for example, where the largest star that forms keeps increasing until
the cloud is destroyed. The uniformity of the IMF over time and space
argues for a process that is independent of specific properties of
molecular clouds. This uniformity, along with the common observation
of young stars in hierarchical clusters that resemble the structures of
interstellar gas, suggests that the IMF, at least in the power law range,
is a by-product of turbulence.  Detailed physical processes may affect
the turnover at low mass, but even this may have a universal character
arising from a combination of turbulence-induced variations in the local
cloud properties, and fragmentation in unstable cloud pieces.
\end{abstract}

to appear in "Star Formation from the Small to the Large Scale,"
eds. F. Favata, A.A. Kaas and A.  Wilson, ESA Pub. Div.,
ESTEC, Noordwijk, Netherlands, ESA SP-445, in press.

\section{Introduction}

The stellar initial mass function (IMF) is a property of star formation
that has recently garnered so much high-quality data that the general
framework for its understanding may soon be at hand. Several relevant
observations are reviewed here. Not all of these observations are
directly related to the IMF; some are more about star formation in
general than about the relative distribution of stellar masses in each
particular region of star formation. Still, when taken as a whole, there
is little alternative but the view that stars virtually freeze out of a
gas that is structured by compressible turbulence, taking with them a
universal signature of the mass distribution enforced by these motions.
We are led to the conclusion that this process happens quickly on a
dynamical time, and over a very wide range of scales, with little or no
feedback and little sensitivity to the gaseous and galactic properties
around it.

There have been two difficult aspects of the IMF problem: understanding
what all of the star formation processes have in common, and sorting
through the selection effects and assumptions that are implicit in each
particular IMF observation. 

For the first problem, the overall shape of the IMF may not depend much
on the detailed processes of star formation: the IMF is an average over
many of these processes and it always looks about the same. There should
be a universal process at work, such as turbulence, that gives the
basic IMF shape, while the detailed processes specific to each region
may only modify the IMF by small amounts.  If this is the case, then we
should be able to understand most of the IMF using only the properties
of turbulence.

The second of these problems is related to the observation that slight
variations in the IMF are still present from region to region, even though
the overall IMF is somewhat uniform (Scalo 1998). Many of these variations
may to be the result of three things: (1) pervasive selection effects,
such as differential aging of high and low mass stars, which leads to
the loss of some fraction at the high mass end when stars form more
or less continuously in a large region for longer than several million
years, or differential drift into the field resulting from the longer
age or higher speeds of low mass stars; (2) mass segregation in clusters
resulting in part from gas and small-star drag on the massive protostars,
and (3) possible shifts in the basic mass scale for star formation in
extreme environments. These systematic variations, along with stochastic
variations from small number statistics in limited surveys, are always
present to some degree in the observations.  We discuss below another
possible variation with the density of the region of star formation.

\section{Star formation in a crossing time}
\label{sect:rapid}

After we thought for a long time that star formation must be slow in
molecular clouds, perhaps to avoid the galactic catastrophe discussed
by Zuckerman and Evans (1974), and after many attempts to explain this
slowness by star-formation feedback (Norman \& Silk 1980; Franco \&
Cox 1983; McKee 1989), magnetic support (Mouschovias 1976; McKee 1989),
and turbulence (Bonazzola et al. 1987; Leorat, Passot, \& Pouquet 1990;
Vazquez-Semadeni \& Gazol 1995), the observations now suggest just the
opposite. Star formation seems to be fast on every scale in which it
occurs, from subparsecs to kiloparsecs (Elmegreen 2000a). Fast means
that the star formation process begins and ends in only a few dynamical
timescales in a cloud. Thus small scales form stars in a short time,
measured in years, and, large scales form stars over a longer time,
but both times are comparable to $\left(G\rho\right)^{-1/2}$ for average
local density $\rho$. Because of the way turbulence structures the gas,
i.e., in fractal or multifractal patterns, bigger scales have smaller
average densities, i.e., more and more of the volume is occupied by low
density gas as the scale increases.

This change in thinking is based on direct and indirect observations.
Direct observations are the age ranges for embedded and young clusters.
The age range for Trapezium stars in Orion is 1 My or less (Prosser et
al. 1994; Palla \& Stahler 1999). In L1641 it is about the same (Hodapp
\& Deane 1993). NGC 1333 has a large number of short-lived jets and
Herbig-Haro objects (Bally et al. 1996), so things are happening quickly
there too. In NGC 6531, the age spread is immeasurably small (Forbes
1996).

These short time scales are all comparable to a few crossing times in
the cloud cores. The average stellar density in the Trapezium cluster is
$\sim10^3$ M$_\odot$ pc$^{-3}$, corresponding to several thousand stars
pc$^{-3}$ (Prosser et al. 1994; McCaughrean \& Stauffer 1994). The
densities are about the same, sometimes a little less, in other young
clusters too (see figure 5 in Testi et al. 1999). Considering that the
efficiency to make a bound or nearly-bound cluster is around 50\% (as
measured in IC 348, for example -- Lada \& Lada 1995), this stellar
density corresponds to an initial H$_2$ density of around
$\sim6\times10^4$ H$_2$ cm$^{-3}$, as is sometimes measured directly
(Lada 1992). The corresponding dynamical time scale is
$\left(G\rho\right)^{-1/2}\sim 0.3$ My.

Palla \& Stahler (1999) suggest that as the Trapezium cloud contracted,
the star formation rate increased, which means that it stayed at a rate
roughly proportional to the instantaneous dynamical time during a factor
of perhaps 10 in density enhancement.

Sometimes star formation occurs in several quick bursts, as in 30 Dor
(Selman et al. 1999), but even then each burst seems to be fast,
spanning a total time of about 2 My per burst in this case. Observations
of longer durations are therefore suspect: the Pleiades cluster has been
claimed to have a prolonged star formation period, perhaps 30 My (Siess
et al. 1997; Belikov et al. 1998), but this could result from a mixture
of multiple events (Bhatt 1989), or even uncertain stellar evolution
times.

The age spread in a whole OB subgroup is generally larger than it is
in any one compact core that might form in the subgroup. Age ranges of
$\sim2$ My seem typical (Blaauw 1964; Massey et al. 1995a). The age range
in a whole OB association, with several subgroups, is larger still,
perhaps 10 My. These larger scales correspond to smaller densities,
however, and to proportionally larger dynamical times. Even larger scales
include star complexes (Efremov 1995), such as Gould's Belt, which
typically take $30-50$ My to finish. Star complexes were originally
discovered by Efremov (1978), and are defined by concentrations of
supergiants and Cepheid variables, which are sensitive to this longer age
range. {\it Each type of star that is used for an observation is associated
with a particular scale for clumping: the longer lived stars highlight
larger regions.}

This correlation between age range and size is not the result of
expansion from a common center, which would make the age range increase
linearly with size. It is not the result of stochastic propagation of
star formation either; then the age range would increase with the square
of the size. In fact, the age range increases with the square root of
the size of the region, as determined from the distribution of Cepheid
variables (Elmegreen \& Efremov 1996) and clusters (Efremov \& Elmegreen
1998) in the LMC. The square root dependence identifies {\it turbulence}
as a controlling factor. Moreover, the constant of proportionality in
the duration-versus-size correlation is about the same as in the
analogous correlation between crossing time and size for molecular
clouds and their clumps (Elmegreen 2000a). Thus star formation on a wide
range of scales, from 20 pc to 1 kpc in the LMC, operates on about 1.5
turbulent crossing times for the associated gas. The result is a
hierarchy of clusters in both age and position: small regions of star
formation come and go, presumably in recycled gas, in the time it takes
the larger region surrounding them to finish (see review in Elmegreen et
al. 1999).

The hierarchical structure of young stars in star-forming regions
provides indirect evidence for {\it relatively} rapid star formation:
if we still see the stars inside a cluster with strong subclustering,
reminiscent of the hierarchical structure in molecular gas, then the
stars cannot have moved very far from their origins. They probably do not
even have time to cross from one side of the cloud core to the other; if
they did, they would mix up and not be hierarchical anymore (Elmegreen
2000a). Hierarchical structure in embedded young clusters is commonly
seen; e.g., in IC 348 (Lada \& Lada 1995), NGC 3603 (Eisenhauer et al.
1998), W33 (Beck et al. 1998), NGC 2264 (Piche 1993), and G 35.20-1.74
(Persi et al. 1997). Elson (1991) found spatial substructure in 18 LMC
clusters, and Strobel (1992) found age substructure in 14 young clusters.

These observations, both direct and indirect, suggest that star formation
is relatively rapid over a wide range of scales. If this is the case, then
{\it there is not enough time for a protostar to move around in a young
cluster and either accrete the ambient gas as it moves or coalesce with
other protostars.} This rules out a large class of models for the IMF.

A good example is provided by the recent IR continuum observations of
young protostars in Ophiuchus and Serpens (Motte, Andr\'e, \& Neri 1998;
Testi \& Sargent 1998). These protostars are very small compared to their
angular separations and are not likely to collide with each other in even
a few crossing times. Also, their distribution is clumpy like the gas;
they should be more scattered if they are randomly orbiting from clump
to clump inside the cloud core.

This result can be quantified by considering the cross section that
would be necessary for an object to collide with another in
one crossing time. If the density of protostars is $n_3\times1000$
pc$^{-3}$, and the radius of the cluster is a typical $R_{cl,0.2}*0.2$
pc (Testi et al. 1999, Fig. 1), then the protostar cross section must be
\begin{equation} \pi R_{protostar}^2=\sim{{\left(10^4 {\rm
AU}\right)^2}\over {n_3R_{cl,0.2}}}. \end{equation} This is such a large
cross section that at typical cluster densities, only binary stars and
disks should be affected by protostar interactions. Older models that
assumed protostars move around for many crossing times in a cloud core
got coalescence with smaller cross sections, but they also had to assume
much higher cluster densities (Silk \& Takahashi 1979; Bastien 1981;
Larson 1990; Zinnecker et al. 1993; Price \& Podsiadlowski 1995;
Bonnell, Bate, \& Zinnecker 1998).

Before we leave this topic, it is worth clarifying why rapid star
formation of the type discussed here, i.e., hierarchical in position and
time, does not lead to a catastrophic starburst in the whole galaxy, as
envisioned by Zuckerman \& Evans (1974). The reason has two observational
sides: On a large scale, the star formation rate in a whole galaxy
is slow because it follows the local dynamical time scale, just like
individual clouds (Elmegreen 1997; Kennicutt 1998). This time scale
is very large, comparable to the orbit time. On a small scale, most
of the CO-emitting gas is not able to form stars: it is either too low
in density as part of an interclump medium inside generally molecular
clouds, or too high in density and transient because of intermittent
turbulent effects. Only a small fraction, like 1\% (McLaughlin 1999),
of the mass in any molecular cloud actively participates in the star
formation process. In this active gas, the efficiency of conversion of
gas into stars is usually high, like 10\% or more. To put it in another
way, even though star formation always operates locally on a dynamical
time scale, the total CO mass in the Galaxy is not turning into stars on
the {\it average} dynamical time scale because the gas is {\it fractal},
i.e., mostly hollow, and the average or excitation density that is used
to determine this average timescale rarely occurs in any local region.

\section{Is the IMF independent of cluster density?}
\label{sect:density}

The inability of stars or protostars to collide directly in even the
densest environments suggests at first that the IMF should be
independent of cluster density. The binary star fraction and relative
disk fraction or disk mass should not be independent of cluster
environment, but the star masses should. This conclusion is right in
some sense, but a generalization of it to all stars might be a bit too
hasty. Cluster density also affects the rate of accretion of ambient gas
onto a protostar, and different densities might lead to different IMFs
because of different accretion rates.

The geometry for the accretion of gas onto a protostar is not really
known. In some models (e.g., Bonnell et al. 1998), a protostar is
assumed to move around in a gaseous medium of uniform, or at least
smooth, density, and to accrete this gas as it moves. Interstellar
clouds are not uniform, however. They seem to be fractal with most of
the mass occupying a small fraction of the volume. If this is the case,
then the model of moving accretion would not build up much stellar mass:
most of the time the moving stars would be in regions with very low
densities.

On the other hand, if the star accretes virtually all of its mass from
a clump, and consequently comoves with that clump because of momentum
conservation, then the stellar mass is more a reflection of the clump
mass, rather than the accretion rate multiplied by a time. We might as
well assume, then, that the star mass is following the clump mass, and
not worry too much about how the accretion actually occurs.  This is the
basic assumption in my recent IMF models (Elmegreen 1997, 1999a, 2000b,c).

There is a third possibility, however. It could also be that by the
time a very dense core forms inside a molecular cloud, the gas is no
longer highly fragmented and fractal. This could be because the Mach
number of the turbulence is relatively low in this case; i.e., the line
width may be nearly thermal (and the temperature may be elevated). If
there is enough mass to form more than one star, then the model for
moving accretion could apply. And in a dense cluster, there could also
be enough time for such thermal cores to coalesce and build up in mass
(because at the Jeans size, these cores would be fairly big). Then we
are faced with the interesting possibility that dense clusters might end
up with a different IMF than low density star-forming regions. That is,
the high mass stars may accrete faster than the low mass stars in such
a uniform environment, and so have a different contribution to the net
IMF than would a low density region. This is an old idea (Larson 1978;
Larson 1982; Zinnecker 1982). What do the observations say about it now?

First of all, the IMF is in fact steeper in low density regions than in
clusters. A compilation of the observations in Elmegreen (1997, 1999a)
demonstrated this. For example, the slope $x$ in the power law part,
written as $M^{-1-x}d\log M$, is in the range from 1.5 to 2 for the
local field stars (Garmany, Conti, \& Chiosi 1982; Scalo 1986; Humphreys
\& McElroy 1984; Blaha \& Humphreys 1989; Basu \& Rana 1992; Kroupa,
Tout, \& Gilmore 1993; Tsujimoto et al. 1997), whereas $x=1.35$ for the
Salpeter function is more appropriate for clusters. The same steep slope
applies to the unclustered young stars in Orion, although the tightly
clustered stars have $x$ in the range from 1 to 1.5 (Ali \& DePoy 1995).
In the whole Orion field, the IMF is steep as well (Brown 1998). J.K.
Hill et al. (1994), and R.S. Hill et al. (1995) also found that LMC
clusters have significantly steeper slopes in regions of low young star
density than high young star density. An even more extreme case is
considered by Massey et al. (1995b), who find that the remote field in
the LMC has $x\sim4$.

The problem with these observations is that they all might contain
selection effects. One example is differential drift, where the low mass
stars, with their longer lives and possibly higher velocity dispersions,
drift further into the field, or into the low density regions, than high
mass stars. Segregation of high mass stars to the potential wells of
young clusters could produce the steepening effect at low density too
(and a relatively shallow IMF in the cluster cores). Such segregation
would have to be rapid, however (Hillenbrand \& Hartmann 1998; Bonnell
\& Davies 1998). Also important might be the failure to correct for the
loss of evolved massive stars in a region that has been forming stars
for a long time. The color magnitude diagram may show only the youngest
high-mass stars, and the oldest could be gone by now. This might be the
case for the IMF in a whole OB association, which could have been
forming stars for a period of 10 My or more, much longer than the
lifetime of the most massive stars. In fact, because the duration of
star formation increases with the size of the region, as shown above,
the IMF might be systematically steeper in the larger, lower density
regions simply because of a lack of corrections for this subtle aging
effect.

On the other hand, the IMFs in dense clusters are surprisingly invariant
for the intermediate-to-high mass range, spanning a factor of several
hundred in cluster star density (Massey \& Hunter 1998; Luhman \& Rieke
1998). Thus, whatever is happening in one cluster must be happening in
all clusters. Moreover, this IMF is very close to the Salpeter IMF,
namely with a slope in the range from -1 to -1.5 on a logarithmic scale
at intermediate to high mass. The same slope applies to galaxies in
general (sect. \ref{sect:gal}).

What complicates this bimodal approach is that the galaxy-wide IMF is
not particularly sensitive to the slope of the IMF at the very highest
masses, except for the requirement that there not be too many high mass
stars, e.g. in the 50-100 M$_\odot$ range, to avoid an unusually high
abundance of oxygen compared to iron (Wang \& Silk 1993). If the
galaxy-wide IMF began to drop at around 50 M$\odot$, and had fewer 100
M$_\odot$ stars than the number expected from an extrapolation of the
Salpeter function, then we probably would not know this. However, dense
clusters like 30 Dor have many O3 stars with masses of around 100
M$_\odot$ and a Salpeter slope out to at least this value
(Massey \& Hunter 1998). {\it Thus it is
conceivable that dense clusters form proportionally more $>50$ M$_\odot$
stars than lower density regions.} It may also be true in this case
that the majority of stars cannot form in dense clusters, because then
the galactic abundance of $>50$ M$_\odot$ stars, and perhaps of oxygen,
would end up too high. Thus there remains a possibility that cluster
density affects the IMF at the very highest masses in ways that are
difficult to observe right now. A discussion of various constraints on
the high mass IMF, including stochastic IMF models that extend into this
range, is in Elmegreen (2000c).

\section{The cluster IMF equals the Galaxy IMF}
\label{sect:gal}

A surprisingly strong constraint on the theory of the IMF comes from the
simple observation, mentioned above, that the cluster IMF slope is about
the same as the galaxy-integrated IMF slope. The latter comes from
observations of emission-line equivalent widths (Kennicutt, Tamblyn \&
Congdon 1994; Bresolin \& Kennicutt 1997), in which the emission line
measures the massive star flux and the continuum measures the low mass
star flux. It also comes from color magnitude diagrams in the LMC and
local dwarf galaxies (Greggio et al. 1993; Marconi et al. 1995; Holtzman
et al. 1997; Grillmair et al. 1998), as well as from the relative
abundances of Fe and O. The latter appear constant in a wide variety of
systems, including QSO damped Ly$\alpha$ lines (Lu et al. 1996) and
Ly$\alpha$ forest lines (Wyse 1998), the intracluster medium (Renzini et
al. 1993; Wyse 1997, 1998), elliptical galaxies (Wyse 1998), and normal
spirals, including the Milky Way.

This apparent equivalence between the cluster and integrated IMFs was
not previously recognized as a constraint on the IMF models because for
a long time it was not known that high mass stars always form along with
low mass stars. Twenty years ago, bimodel star formation models proposed
that high and low mass stars may form in different regions. These models
are no longer popular, however (see review sections in Elmegreen 1997,
1999a). For the more realistic case in which high mass clouds make both
high mass and low mass stars, giving a normal IMF, and for the common
observation that low mass clouds make primarily low mass stars, we would
have the unusual circumstance that far more low mass stars should be
forming than high mass stars compared to the normal IMF were it not
required that stars of all masses form randomly in clouds of all masses.
That is, if high mass clouds make both high mass and low mass stars, but
low mass clouds make only low mass stars, then the large number of low
mass clouds compared to high mass clouds would give a summed IMF that is
steeper than the IMF in each region.

Instead, it must be true that even low mass clouds occasionally form
intermediate or high mass stars, although we rarely see this because of
sampling effects; i.e. it takes a lot of stars to get an IMF
sufficiently populated to form a high mass star. To put it differently,
the observations seem to require that ten low mass clouds have the same
probability of forming a high mass star as a single cloud with ten times
the mass. In retrospect, considering the fractal structure of clouds,
this is neither surprising nor unobserved. When viewed from a distance,
small clouds are seen to be parts of large clouds, so when we see a
high-mass star forming in an extended region of star formation with a
large total cloud mass, closer examination should show that this star is
really forming in a smaller subcloud, and that other smaller subclouds
may have not have such massive stars at all. Only when we lose the
multi-scale perspective, by studying only the local regions of star
formation for example, do we have difficulties understanding how small
clouds can make large stars.  This statement is independent of the
peculiarities of high mass star formation, which may involve hot cores
or dense clusters, unlike low mass stars; it is only about sampling
where such peculiarities are likely to occur among all of the clouds in
a region.

We can also see how this constraint on unrestricted starbirth
locations follows from the observations by considering a simple
theoretical model (Elmegreen 1999a). Suppose that the cloud or
cluster mass spectrum is \begin{equation}N(M_{cl})dM_{cl}\propto
M_{cl}^{-\gamma}dM_{cl}\end{equation} and that the largest star
mass that forms in a cluster of mass $M_{cl}$ is $M_{L}\propto
M_{cl}^{\alpha}$. Suppose also that the IMF in each region is
\begin{equation}n(M) dM = n_0 M^{-1-x} dM\end{equation} out to the
largest star mass, $M_L$. Setting $N(M_L)dM_L= N(M_{cl})dM_{cl}$, we
can convert the individual IMFs into a summed IMF with the equation:
\begin{equation} n_{gal}(M)= \int_M^\infty n(M|M_L)P(M_L)dM_L \propto
M^{-1-x_{eff}} \end{equation} where $n(M|M_L)$ is the conditional
probability of forming a star of mass $M$ in a region with a maximum
mass $M_L$. The integration gives $x_{eff}=\left(\gamma-1\right)/\alpha$,
which is remarkably independent of the local IMF slope, $x$.

Now, $\gamma\sim2$ for clusters and probably for clouds also,
considering the hierarchical structure (Fleck 1996; Elmegreen \&
Falgarone 1996; Elmegreen \& Efremov 1997). Then the constraint that the
local IMF equals the global IMF means that $x_{eff}=x$, which becomes
$x=1/\alpha$. That is, the largest star mass must increase with the
cluster mass as $M_L\propto M_{cl}^{x}$. This is in fact observed
(Elmegreen 1983), but it also follows theoretically from purely random
sampling with all stars equally likely to form in all clouds. The reason
is that the largest mass star comes from the IMF through the equation
\begin{equation} \int_{M_L}^\infty n(M)dM=1, \nonumber\end{equation}
which implies that $n_0=xM_L^x$. The total cluster mass in the power law
range is, similarly, \begin{equation} M_{cl}=\int_{M_{smallest}}^\infty
Mn(M)dM ={{x}\over{x-1}}M_L^xM_{smallest}^{1-x} \end{equation} for
smallest mass star $M_{smallest}$ in the power law range. These two
equations give $M_L\propto M_{cl}^x$.

Thus {\it large mass clouds are more likely to form high mass stars, but
only because they form more stars overall.} This is the implication of
the observation that the cluster IMF equals the galaxy-wide IMF.

As a result of this implication, we can also state that star mass does
not increase monotonically with time in a cluster as more and more stars
form, because of some gaseous heating or heightened turbulence for
example. This rules out a large class of sequential IMF models. If a
model has this character, it may explain the IMF in any one cloud, but
it cannot explain the IMF in the sum of all clouds.

There may be an interesting exception to this equality between
the cluster IMF and the galaxy IMF in the observation by Massey et
al. (1995b) that the slopes of the IMFs in the extreme field regions
of the LMC and Milky Way are much steeper than the slopes in clusters.
This goes in the same direction as the density dependence discussed
in Section \ref{sect:density}, so the result may have something to do
with protostellar accretion, but there is also a simpler explanation:
In the extreme field, the pressure and cloud density are likely to be low
and the overall star formation rate low. Then OB stars may more readily
disrupt their clouds and halt further star formation once they appear. If
we consider the disruptive power of Lyman Continuum radiation and how
it scales with cloud and cluster mass, then the value of $\alpha$ used
above in the expression equals about 0.25 (Elmegreen 1999a).  This gives
$x_{eff}=4$, which is what Massey et al. (1995b) observe. Elsewhere,
$\alpha\sim1/x$ and $x_{eff}=x\sim1.35$, so massive stars should not
readily halt star formation in their clouds. Presumably this is because
star formation is usually too fast.

\section{Young star fields are fractal, like the gas}

The final clue to the origin of the IMF mentioned in the introduction
is the observation that young star fields are hierarchically clumped,
or fractal, just like the gas. Reviews of this property are in
Elmegreen \& Efremov (2000) and Elmegreen et al. (1999). Sometimes the 
hierarchy can be traced for 5 levels, from the scale of a giant patch of
star formation in a spiral arm to subclumping inside a compact cluster
(e.g., the cases of W3 and Orion are described in detail in Elmegreen
et al. 1999). Various observations of subclustering of stars inside
clusters were already mentioned above.

The gas has a similar structure, as is well known from fractal
cloud studies (e.g., Scalo 1985; Falgarone et al. 1991; Stutzki
et al. 1998). The mass spectrum of clouds probably comes from this
structure too (Elmegreen \& Falgarone 1996). The origin of all this
fractal structure is presumably turbulence. Computer models reproduce
it as well as can be expected at the present time (MacLow et al. 1997;
Elmegreen 1999b).

The implication is that the stars freeze out of turbulent gas rather
quickly, without moving much from their formation sites. Moreover, if
they form in turbulence-generated clumps, then they probably have masses
that are proportional to the turbulence-generated masses, i.e., to the
local clump masses in a turbulent, self-gravitating medium. Models for
this process are in Elmegreen (1993) and Klessen et al. (2000). In that
case the origin of the IMF power law, or, if not an exact power law,
then something close to a power law, such as a log-normal distribution,
is a natural consequence of turbulence far removed from any boundaries.
Turbulence produces power law velocities and scale-free structures, like
fractals or multi-fractals (Sreenivasan 1991). The questions is, how
exactly does turbulence make the IMF? We do not know yet, but we can
come close with a simple model that has all the essential physics.

\section{A model for the IMF in turbulent clouds}

An interstellar cloud is somewhat self-similar over a wide range of
scales. The thermal Jeans mass hardly shows up in the correlations
between total linewidth and size, or density and size. For this reason,
the formation processes of stars are probably self-similar too for a
wide range of masses, at least in the power-law part of the IMF. This
means that in a hierarchical cloud, the mass that goes into a star can
come from any level in the hierarchy, provided the corresponding clump
is sufficiently self-gravitating at some time in its life to make a
star.

The total mass range for clumpy structure in clouds is $\sim10^{10}$,
whereas the mass range for stars in the power law part of the IMF is
only $\sim100$. Because of this large difference in mass range, stars
have to come from only the part of the cloud hierarchy that is fairly
close to the thermal Jeans mass at the total cloud pressure. Below that,
the gas cannot collapse easily. The break in the IMF from the power law
part at intermediate to high mass and the relatively flat part at low
mass occurs at this thermal Jeans mass, which is $\sim0.3$ M$_\odot$
for typical conditions (Larson 1992; Elmegreen 1997). If there is
no preference for scale above the thermal Jeans mass, then stars are
essentially coming from a more-or-less random sample of clumps in the
hierarchical gas distribution. Random samples from a hierarchy produce
an $M^{-2}dM$ mass distribution (Fleck 1996), which is already close
to the Salpeter function of $M^{-2.35}dM$. Random sampling from Fourier
$k$-space produces a $k^2dk=M^{-2}dM$ spectrum too (Elmegreen 1993). This
is a good start to look for a theory of the IMF.

The next step is to recognize that the sampling process cannot be
completely uniform on all scales. As we have seen, dynamical events work
faster at higher densities. In a fractal cloud, the smallest fragments
have the highest densities, so we should really be considering a random
sample with a rate proportional to the dynamical rate, which scales as
the square root of the local density. In this case, smaller mass clumps
form stars more often than higher mass clumps, and that steepens the IMF
from a slope of 2 to about 2.15 (Elmegreen 1997). The same steepening
occurs if we think of the turbulent rate as a function of wavenumber $k$
in phase space too, because the turbulent rate varies approximately as
$k^{1/2}\propto M^{-1/6}$, which steepens the IMF for purely turbulent
sampling to $k^{5/2}dk\propto M^{-13/6}dM$ (Elmegreen 1993).

Now there are two additional effects, one that takes care of itself
automatically in the above picture, and another that is likely to happen
anyway, and which requires a bit more physics. The first effect is a
competition for gas: when a dense, low-mass clump turns into a star,
the gas that made it is no longer available to make another star. This
effect steepens the IMF to a slope equal to about the Salpeter value,
$2.3$ (or 1.3 for intervals of $\log M$).  The reason is that each mass
structure surrounding the first star in the hierarchy of structures has
a little less gas to make its second star.

The other effect is that structures that are very large will evolve so
slowly that the gas inside of them should turbulently remix and make new
dense cores before the larger scale itself can make a star. This differs
from the competition for mass above, which works even if the turbulent
structures are static. With the second effect, the gas that is left over
after the formation of a low mass star in an original low mass core can
get recycled by turbulent motions into forming more low mass cores, and
this may form more low mass stars without ever getting into a large mass
star on the larger scale. This constraint involves timing. Any scale
that has a star formation time much larger than the turbulent mixing
time of the smallest star-forming scale inside of it is not likely to
form a single large star, but will form numerous smaller stars instead.

This second effect has been simulated in a computer model (Elmegreen
2000c) by rejecting any previously chosen clump of mass $M$ with a
probability of $1-e^{-t(M)/t(M_J)}$ for crossing time $t(M)$ on scale
$M$ and for minimum mass $M_J$, which is the Jeans mass or some other
minimum. If $M\approx M_J$ and $t(M)\approx t(M_J)$, then stars freeze out
at scales equal to or less than $M$ without much turbulent remixing. If
$t(M)>t(M_J)$, then even after all the initial low mass clumps turn
into stars, more low mass clumps will still have time to form from
the residual gas and turn into more low mass stars. Then nothing is
left over for a high mass star of mass $M$. This ratio $t(M)/t(M_J)$
is the average number of turbulent crossings for scale $M_J$ that occur
during a turbulent crossing time at scale $M$. It is the mean waiting
time for significant mixing on scale $M$. Thus $e^{-t(M)/t(M_J)}$ is
the Poisson probability that no significant remixing occurs.  The clump
mass is related to the crossing time by $M/M_J\propto [t(M)/t(M_J)]^5$.
The observed IMF requires that $t(M)/t(M_J)$ be no more than order
unity, perhaps at most 2, limiting $M/M_J$ to less than several hundred
before the fall off in the high mass IMF becomes steep.  This limit in
$t(M)/t(M_J)$ is consistent with the requirement of rapid star formation,
discussed in Section \ref{sect:rapid}.

Model IMFs with this timing constraint are shown in Elmegreen (2000c).
The slope at intermediate mass steepens a little, from $2.3$ to $2.35$,
which is in fact the Salpeter value, and it steepens a lot above $\sim100$
M$_\odot$, making the formation of extremely massive stars as unlikely
as the observations require.

\section{Conclusions}

The processes of star formation are not yet well enough understood to
trace in detail the sequence of events that differentiates high and low
mass stars. There is so much uniformity in the IMFs from different
regions, however, that many of these factors may not be important for
the final distribution anyway. Somehow the averaging inherent in
plotting a histogram of stellar masses erases the memory of the
different physical processes that were involved in forming the stars.
Triggered regions of star formation have about the same IMFs as
quiescent regions; large region have about the same functions as small
regions, aside from sampling statistics, and moderately old regions are
about the same as the youngest. What this means is that stars with the
same mass could have formed in very different environments, even with
different processes, but we would not necessarily know this from a mere
histogram of final star mass. The averaging has erased the details.

The previous sections have proposed that the approximately power law part
of the IMF is somehow tracing the power-law conditions in star-forming
clouds that are continuously established by pervasive turbulence. If true,
then we do not need a theory of star formation to explain the IMF, but
rather a theory of turbulence. This is the reverse of what most studies
have been after: previous theories of the IMF began with a model for how
stars form, usually ignoring the turbulent properties of the gas that goes
into these stars, and then sampled the parameter space or the competitive
processes until the final stellar masses were obtained. Now we suspect
that {\it any} reasonable theory of star formation can give the same
final IMF, even many of the theories that have already been proposed
on physical grounds, provided only that they operate in a fractal,
hierarchically-structured and turbulent medium.  This may be why random
sampling from hierarchical clouds at a rate that scales with the square
root of the local density gives the right result: all theories of star
formation have this basic scaling for the rate at which things happen.

The flattening at low mass down to the brown dwarf state may be a
different matter. Here there is a characteristic mass, the mass at the
lower limit to the Salpeter power law, and there may be a different
reason for the mass distribution function below this limit than above.
At the most fundamental level, however, this low-mass distribution is
not {\it that} different from the high mass distribution: both are power
laws for a factor of $\sim100$ in mass range; they just have different
powers. Maybe some scale-free properties of cloud or collapse dynamics
are involved at low mass too. Regardless, the boundary itself has to
contain some other physics to get the mass scale. For this reason, the
mass at the boundary between the high-mass power law and the low-mass
flat part may be expected to vary over the extreme range of star-forming
environments. Such variations have been suggested for starburst regions,
but direct observations at the boundary mass are still lacking.

The previous sections also emphasized the importance for the IMF of the
dynamically rapid timescale for star formation in most regions. This
seems to rule out a class of models that depends strongly on protostellar
orbits in the cloud or on protostellar interactions. The observation of
a similar IMF in clusters and in whole galaxies suggests further that
stars of all masses can form in clouds of all masses. The issue here
is that clouds are basically fractal in structure, so clouds of all
masses are contained within, or are parts of, clouds with all higher
masses. The definition of a cloud mass is vague. One uncertainty is
whether dense clusters really have different high-mass IMFs than the
average for all stars. Dense clusters may, in fact, have a slight excess
of high-mass stars compared to the average for all stars; it would be
very difficult to know this at the present time. If true, then accretion
and coalescence effects may be important in dense cluster cores, but
the fraction of stars which form under these conditions would have to be
low. The Salpeter function found in dense clusters out to 100 M$_\odot$
or more cannot continue indefinitely. If it did, then a galaxy the size
of ours would have a few 1000 M$_\odot$ stars (at birth), just by sampling
the IMF for a large total mass.

Progress in understanding the IMF should come from two fronts:
observations of statistically significant IMFs in a variety of different
environments, sampling extremes in density, temperature, and pressure, and
computational modeling of mass segregation processes in self-gravitating,
turbulent, magnetic gas. Further IMF modeling in whole galaxies and
careful studies of the timescale for star formation would be useful too.

\end{document}